\documentstyle[aps,prd]{revtex}

\begin{document}

\title{Inhomogeneous Neutrino Degeneracy and Big Bang Nucleosynthesis}
\author{Scott E. Whitmire,$^{1}$ and Robert J. Scherrer$^{2,3}$}
\address{$^1$Department of Physics, State University of New York,
Buffalo, NY~~14260}
\address{$^2$Department of Physics, The Ohio State University,
Columbus, OH~~43210}
\address{$^3$Department of Astronomy, The Ohio State University,
Columbus, OH~~43210}

\maketitle
%

\renewcommand{\baselinestretch}{1.3}
\begin{abstract}

We examine Big Bang nucleosynthesis (BBN) in the case of inhomogenous
neutrino degeneracy, in the limit where the fluctuations
are sufficiently small on large length scales
that the present-day element abundances are homogeneous.
We consider two representive cases:  degeneracy of the electron neutrino
alone, and equal chemical potentials for all three neutrinos.
We use a linear programming method to constrain an arbitrary
distribution of the chemical potentials.
For
the current set of (highly-restrictive) limits on the primordial
element abundances, homogeneous neutrino degeneracy barely
changes the allowed range of the baryon-to-photon ratio $\eta$.
Inhomogeneous degeneracy allows for little change in the lower
bound on $\eta$, but the upper bound in this case
can be as large as $\eta = 1.1 \times 10^{-8}$ (only $\nu_e$
degeneracy) or $\eta = 1.0 \times 10^{-9}$ (equal degeneracies
for all three neutrinos).
For the case of inhomogeneous neutrino degeneracy, we show that there
is no BBN upper bound on the neutrino energy density,
which is bounded in this case only by limits from structure formation
and the cosmic microwave background.

\end{abstract}
\pacs{PACS numbers: }
%

\twocolumn

\section{Introduction}

Although the standard model of Big Bang nucleosynthesis (BBN) is highly
successful (for a recent discussion, see references \cite{burles,OSW})
many variations on this model
have been proposed \cite{malaney}.  One of the most frequently
investigated variations on the standard model is neutrino
degeneracy, in which each type of neutrino is allowed
to have a non-zero chemical potential
\cite{wagoner} - \cite{kawasaki}, and a number of recent models
have been proposed to produce a large lepton degeneracy
\cite{gelmini} - \cite{murayama}.

More recently, Dolgov and Pagel have suggested the possibility
of inhomogeneous neutrino degeneracy \cite{dolgov}.  Their
model was proposed to explain the apparent discrepancy between
various measurements of the primordial deuterium abundance
in high-redshift Lyman-alpha clouds.  Here we consider a more
mundane possibility:  that the neutrino chemical potential
is inhomogeneous, but on much smaller scales.  In particular,
we assume that the amplitude of the inhomogeneities is small
on length scales larger than the typical baryonic diffusion
scales after nucleosynthesis, so that the element
abundances are homogeneous today.  We calculate the
element abundances for this scenario and compare to
observational limits.  Using a method similar to that
in reference \cite{leonard}, we can simulate arbitrary
distributions of the neutrino chemical potential, and
so determine the upper and lower bounds on the baryon-to-photon
ratio $\eta$ in this model.

In the next section, we discuss our model for inhomogeneous
neutrino degeneracy and its physical consequences.  In
Section 3, we use a linear programming technique
to calculate upper and lower bounds on $\eta$ in
this model.  Our conclusions are summarized in
Section 4.  We find that when the chemical potential
is inhomogeneous, there are no BBN limits on the overall
neutrino energy density; the only limits in this case come
from other cosmological considerations such as structure
formation \cite{kang,freese} or the CMB \cite{cmb}.  Not surprisingly,
inhomogeneous neutrino degeneracy allows for a wider range of
values for $\eta$ than does homogeneous degeneracy.

\section{Inhomogeneous Neutrino Degeneracy}

Consider first the case of homogeneous neutrino degeneracy.
In this case, each type of neutrino is characterized
by a chemical potential $\mu_i$ ($i = e, ~\mu, ~\tau$),
which redshifts as the temperature, so it is convenient
to define the constant quantity $\xi_i \equiv \mu_i/T_i$.
In terms of $\xi_i$, the neutrino and antineutrino number densities
are given by
\begin{equation}
\label{nnu1}
\nu_i = {1\over 2 \pi^2} T_\nu^3 \int_0^\infty {x^2 dx \over 1+
\exp(x - \xi_i)},
\end{equation}
and
\begin{equation}
\label{nnu2}
\bar \nu_i = {1\over 2 \pi^2} T_{\bar \nu}^3 \int_0^\infty {x^2 dx\over 1+
\exp(x + \xi_i)},
\end{equation}
while
the total
energy density of the neutrinos and antineutrinos is
\begin{equation}
\label{rhonu}
\rho = {1\over 2 \pi^2} T_\nu^4 \int_0^\infty {x^3 dx\over 1+
\exp(x - \xi_i)}
+ {1\over 2 \pi^2} T_{\bar \nu}^4 \int_0^\infty {x^3 dx\over 1+
\exp(x + \xi_i)}.
\end{equation}
Degeneracy of the electron neutrinos alters the $n \leftrightarrow p$
weak rates relevant for BBN through the number densities
given in equations (\ref{nnu1}) and (\ref{nnu2}),
while the change in the expansion
rate due to the altered density in equation (\ref{rhonu}) affects BBN
for degeneracy of any of the three types of neutrinos 
(see, for example, reference \cite{kang}
for a more detailed discussion).

What happens if this degeneracy is not homogeneous, as assumed
in almost all previous work, but instead
varies with position?  Dolgov and Pagel \cite{dolgov} considered
such a model in order to explain the discrepancy in
observed deuterium abundances at high-redshift.  In their
model, $\xi$ varies on scales $\sim 100 - 1000$ Mpc,
producing an observable inhomogeneity in the present-day
element abundances.  We make the opposite assumption:
we take the variation in $\xi$ to be small
on such large scales, and large on much smaller scales,
so that elements are well-mixed before the present
day, erasing any detectable inhomogeneities.

Although models have been proposed which produce
inhomogeneities in $\xi$ (see, for example, the
discussion in reference \cite{dolgreview}), we will follow the example
of reference \cite{dolgov} and
keep our discussion as general as possible.
In general, one would expect a distribution of fluctuations
in $\xi$ over all length scales.  However, since the neutrinos
are relativistic, they will free-stream and erase any fluctuations
on length scales smaller than the horizon at any given time.
We will make only two assumptions concerning the fluctuations
in $\xi$:  that the fluctuations are significant on large enough
scales to avoid being erased by free-streaming, and that
they become negligible on small enough scales that the
resulting element distribution is homogenous today.

The first of these conditions requires that
the fluctuations are significant
on scales larger than the horizon scale when the $n \leftrightarrow p$
reactions freeze out at $T \sim 1$ MeV.  If this were not
the case, then free-streaming would erase all of the fluctuations
in $\xi$ before BBN began.  This horizon scale corresponds
to a comoving length scale $\sim 100$ pc today.

The condition that the element distribution be homogeneous
today requires
that the fluctuations in $\xi$ decrease sufficiently quickly
with length scale that they have no
significant effect on nucleosynthesis on scales above the element
diffusion length.  Although no detailed studies of
element diffusion have been performed in
connection with inhomogeneous BBN scenarios, it seems safe
to assume that complete mixing of the primordial elements
will occur on scales well within the nonlinear regime today,
$< 1$ Mpc.  By requiring the fluctuations
in $\xi$ to be negligible above this scale, we can also ignore
any constraints from CMB observations, which severely constrain
models with fluctuations on larger scales \cite{dolgov}.

Given these assumptions, we can assume that
BBN takes place in separate horizon volumes, with
the value of $\xi$ being homogeneous within each volume.
At late times, the elements produced within each volume
mix uniformly to produce the observed element abundances
today.

Note that with this set of assumptions, it is no longer
meaningful to talk about the value of $\xi$ for
the neutrinos at the present.  Since the neutrinos
from different horizon volumes diffuse freely up to
the scale of the horizon, the different thermal distributions
with different values of $\xi$ will combine to give a highly
non-thermal neutrino distribution at late times.  Thus, in the inhomogeneous
scenario, it is still possible to put constraints on
the present value of $\rho_\nu$, but it is meaningless
to discuss limits on $\xi$, since the present neutrino
distribution will be non-thermal and cannot be characterized
by a single value of $\xi$.  This effect is present
at some level even at the time of nucleosynthesis.
The neutrinos remain in thermal equilibrium down to
a temperature $T \sim 2-3$ MeV, so that the
neutrino distribution remains thermal down
to this temperature, with a single unique value of $\xi$
in each horizon volume.  However, the $n \leftrightarrow p$
rates do not freeze out until $T \sim 1$ MeV, so that
neutrino free-streaming after decoupling will tend
to produce a somewhat non-thermal background as
early as the beginning of nucleosynthesis.  We have
neglected this effect, which will be negligible in any case if the
neutrino inhomogeneities are confined to comoving scales $> 100$ pc.

We also assume for simplicity that $\eta$ remains uniform
in the presence of an inhomogeneous lepton distribution.
This need not be the case if, for example, baryogenesis
is related in some way to the lepton number \cite{mcdonald,murayama}.

\section{The Effect on Big Bang Nucleosynthesis}

Given our discussion in the previous section, we assume
that the distribution of each type of neutrino is homogeneous within
a given horizon volume during nucleosynthesis,
and characterized by a single degeneracy parameter $\xi_i$ ($i
= e, ~\mu, ~\tau$).
Different horizon volumes may have different values
of $\xi_i$, so we characterize the distribution of
$\xi_i$ by a distribution function $f(\xi_i)$,
which gives the probability that a given horizon
volume has a value of $\xi_i$ between $\xi_i$
and $\xi_i + d\xi_i$.

What form should we choose for $f(\xi_i)$?  In analogy
with the distribution of primordial
density perturbations (and in accordance
with the central limit theorem) the most obvious choice
is a Gaussian distribution.  However, we can consider
a more general case than this.  Using linear programming techniques
like those in reference \cite{leonard}, it is possible to
analyze the general case of an {\it arbitrary} distribution $f$.
Consider first the case where only $\nu_e$ is degenerate, and
suppose that $f(\xi_e)$ is an arbitrary distribution.
Then all of the element abundances will be functions
of $\xi_e$ (for fixed $\eta$), and we can write, for a given
nuclide $A$,
\begin{equation}
\label{XA}
\bar X_A = \int_{-\infty}^\infty X_A(\xi_e)f(\xi_e) d\xi_e,
\end{equation}
where $X_A(\xi_e)$ is the mass fraction of $A$ as
a function of $\xi_e$, and
$\bar X_A$ is the mass fraction of $A$ averaged
over all space; after the matter is thoroughly mixed, $\bar X_A$
will be the final observed primordial element abundance.

In order to test all possible distribution functions $f$,
we can divide the range in $\xi_e$ into discrete bins (not
necessarily all of the same size), and
approximate the integral in equation (\ref{XA}) as a sum:
\begin{equation}
\bar X_A = \sum_j X_{Aj} f_j \Delta \xi_{ej},
\end{equation}
where the dependence of $X_A$ and $f$ on $\xi_e$ is expressed through
their dependence on the bin number $j$.

For each of the elements of interest ($^4$He, D, and $^7$Li)
we have an upper and a lower observational bound.  Thus, for each of these three
elements, we can write down equations of the form:
\begin{equation}
\label{constrain1}
X_{\rm lower~bound} < \sum_j X_{Aj} f_j \Delta \xi_{ej} < X_{\rm upper~bound}.
\end{equation}\
Furthermore, $f(\xi_e)$ is normalized to unity, so
\begin{equation}
\label{constrain2}
\sum_j f_j \Delta \xi_{ej} = 1.
\end{equation}
If we now define 
\begin{equation}
p_j \equiv f_j \Delta \xi_{ej},
\end{equation}
then equations (\ref{constrain1}) and (\ref{constrain2}) become:
\begin{equation}
\label{constrain3}
X_{\rm lower~bound} < \sum_j X_{Aj} p_j < X_{\rm upper~ bound},
\end{equation}
and
\begin{equation}
\label{constrain4}
\sum_j p_j = 1.
\end{equation}
If we put an upper and lower cutoff on these sums, so that
we retain only a finite number of terms, then
equations (\ref{constrain3}) and (\ref{constrain4}) are in the
form of the constraint equations
in a linear programming problem, with the N independent
variables being the $p_j$'s.  In reference \cite{leonard},
the variable under consideration was $\eta$ rather than $\xi$,
so that the final quantity which needed to be maximized
or minimized was the mean value of $\eta$.  In our
case, we wish to determine, for a given value of $\eta$,
whether there is a solution to equations (\ref{constrain3})
and (\ref{constrain4}).  Since there are non-BBN limits on $\rho_\nu$,
we have chosen to take the quantity $\rho_\nu^\prime/\rho_\nu$
as our objective function, where $\rho_\nu^\prime$ is the final
mean total neutrino density in the degenerate case, and $\rho_\nu$
is the neutrino density in the absence of degeneracy.  (These densities
include all three neutrinos and antineutrinos).
We then determine whether a solution exists to our constraint
equations for a given value of $\eta$, and scan through the allowed
range of $\eta$ until we reach an upper and lower value of $\eta$ for
which a solution no longer exists.  At these limiting values for $\eta$,
our linear programming routine gives the minimum possible value
of $\rho_\nu^\prime/\rho_\nu$, which we can compare to other constraints.

We consider two representative cases of interest:  first, the case
where $\xi_e \ne 0$ and $\xi_\mu = \xi_\tau = 0$, which
is equivalent to $\xi_e \gg \xi_\mu, \xi_\tau$, and
the case $\xi_e = \xi_\mu = \xi_\tau$.  The latter is
probably the most physically realistic case \cite{OSTW,mcdonald}.
Although we have discussed our linear programming
procedure only for the case of $\nu_e$ degeneracy, it generalizes
in an obvious way for the case where $\xi_e = \xi_\mu = \xi_\tau$.
We have not considered the most general possible case, in
which all three degeneracy parameters vary independently.
However, as we shall see, arbitrary inhomogeneity in $\xi_e$ alone allows
absurdly large values of $\eta$ to be compatible with BBN,
so there is nothing further to be gained in considering the most general case.

We use for our limits on the element abundances the values
in the recent review in reference \cite{OSW}.  For the
primordial helium-4 mass fraction, ${\rm Y}_P$, we take
\begin{equation}
\label{Y}
0.228 \le {\rm Y}_P \le 0.248.
\end{equation}
The limits on the number ratios of deuterium and lithium-7 to
hydrogen are:
\begin{equation}
2.9 \times 10^{-5} \le {\rm D/H} \le 4.0 \times 10^{-5},
\end{equation}
and
\begin{equation}
1.3 \times 10^{-10} \le {\rm ^7Li/H} \le 2.0 \times 10^{-10}.
\end{equation}
However, a BBN calculation with these limits alone yields no
single value of $\eta$ consistent with all three
sets of limits.  One can argue either that the theoretical uncertainties
are large enough to account for this discrepancy \cite{OSW},
or that one of these sets of limits (most likely lithium) does
not represent the true primordial abundance \cite{burles}.
We have chosen the former approach.  Folding in the theoretical
uncertainties in the BBN predictions from reference \cite{OSW}, we take the following
limits on D/H and $^7$Li/H:
\begin{equation}
\label{D}
2 \times 10^{-5} \le {\rm D/H} \le 5 \times 10^{-5},
\end{equation}
and
\begin{equation}
\label{Li}
1 \times 10^{-10} \le {\rm ^7Li/H} \le 4 \times 10^{-10}.
\end{equation}
We have ignored the theoretical uncertainty in helium-4 because
it represents a much smaller fractional change in Y$_P$.

We wish to emphasize that our general results are fairly
insensitive to small changes in the limits quoted above.
Since we are exploring a rather radical
change to the standard model, we make no effort to perform
an ultra-high-precision calculation.

We used the procedure discussed above to determine the largest
and smallest values of $\eta$ which are consistent with the
limits on Y$_P$, D/H, and $^7$Li/H in equations (\ref{Y}),
(\ref{D}), and (\ref{Li}).  Our mixing procedure
requires the use of mass fractions, rather than ratios to
hydrogen, so we have made this conversion in our calculation.

Consider first the ``standard model" with no degeneracy.
For the limits quoted above, we obtain bounds on $\eta$
of $3.7 \times 10^{-10} \le \eta \le 5.3 \times 10^{-10}$.
Now what happens if we add a homogeneous neutrino degeneracy?
We have calculated the bounds on $\eta$ for the case
in which $\xi_e$ (only) can have an arbitrary value,
and for the case where $\xi_e = \xi_\mu = \xi_\tau$
can be set to any desired value.  For both cases,
we find
that the bounds on $\eta$ are almost unchanged.
(The lower and upper limits are both enlarged by less than
2\%).
While this
might seem surprising in light of earlier similar calculations
\cite{kang},
it is a consequence of the increasingly
narrower limits on the primordial element abundances.
With such sharp limits as those considered here, even
a free variation in $\xi_e$  or in $\xi_e = \xi_\mu  = \xi_\tau$
cannot significantly alter
the limits on $\eta$.
(We could obtain a larger range in $\eta$ by allowing
$\nu_e$ and either $\nu_\mu$ or $\nu_\tau$ to vary independently,
but we would still expect
a narrower allowed range than in reference \cite{kang}
because of the improved observational limits).

Now we proceed to the case of inhomogeneous degeneracy.
As we have noted previously, there is no well-defined
mean final value of $\xi_i$ in this case, since the neutrinos
mix at late times to produce a non-thermal distribution.
However, the mean final value of $\rho_\nu$ is still well-defined,
so we can attempt to constrain it with BBN.
Consider first the case of $\xi_e \gg \xi_\mu, \xi_\tau$.
In this case, all of the element abundances go to zero
in the limit of large $\xi_e$.  Thus, if we take $f(\xi_e)$
to have the form $f(\xi_e=0) \approx 1$, and $f(\xi_e = \xi_0) = f_0 \ll 1$,
where $\xi_0$ is a sufficiently large value of $\xi_e$ such
that all of the element abundances are neglible, then as
we take the limit where $\xi_0 \rightarrow \infty$, the element
abundances approach their values in the standard nondegenerate
model, while $\rho_{\nu_e}$ goes to infinity.  Thus, in the
case of inhomogeneous $\nu_e$ degeneracy, there is no BBN
limit on $\rho_{\nu_e}$.  Of course, there are other cosmological
limits on $\rho_\nu$ in this case, from the requirement
that structure formation not be disrupted by the extra
radiation \cite{kang,freese} and that the extra radiation
not distort the CMB fluctuation spectrum \cite{cmb}.  Our
argument also applies to the case where all three neutrinos
have equal chemical potentials.

There are still interesting limits to be placed on $\eta$.
To determine these limits, we calculated the BBN element
abundances for a grid of values of $\xi$.  We
took $\xi$ in steps of $\Delta \xi = 1.0$ between $\xi = -60$ and
$\xi = 10$.
We embedded a smaller grid between $\xi = -1.0$ and $\xi = 1.0$
in steps of $\Delta \xi = 0.05$.  In calculating the element
abundances for the degenerate case, we used the approximation
given in reference \cite{kang} for the decrease in the neutrino
temperature at large $\xi$.  Although rough, this approximation
is adequate for our purposes.

For the case $\xi_e \gg \xi_\mu, \xi_\tau$, we find
acceptable solutions for $\eta$ in the range
\begin{equation}
3.0 \times 10^{-10} \le \eta \le 1.1 \times 10^{-8},
\end{equation}
while for the case $\xi_e = \xi_\mu = \xi_\tau$,
we have
\begin{equation}
3.1 \times 10^{-10} \le \eta \le 1.0 \times 10^{-9}.
\end{equation}
The actual $\xi$ values of the non-zero bins, along
with the corresponding values for $p_j$ and $\rho_\nu^\prime/\rho_\nu$
are given in tables I and II.  Note that our linear programming
method will always yield a final optimal distribution
for the $p_j$'s in which at most seven of the bins are non-zero
(since equations (\ref{constrain3}) and (\ref{constrain4})
correspond to a total of seven constraint equations); effectively,
this corresponds to a final distribution for $f(\xi)$ which
is a sum of at most seven delta functions (see references
\cite{leonard,NR} for a more detailed discussion).

We see that allowing for a free distribution of the degeneracies
significantly increases the upper bound on $\eta$,
particularly for the case of $\xi_e \gg \xi_\mu, \xi_\tau$,
but
decreases the lower bound only slightly.
Furthermore, the minimum increase in the neutrino density
needed to achieve these lower bounds is inconsistent with
both structure formation considerations \cite{kang} and
CMB observations \cite{cmb}.  On the other hand, the value
of $\rho_\nu^\prime /\rho_\nu$ needed to achieve the upper
bounds on $\eta$ is well within the regime allowed
by both structure formation and the CMB.

\section{Discussion}

Our results indicate that, not surprisingly, the introduction
of inhomogeneous neutrino degeneracy allows for a much wider
range of $\eta$ within the constraints of BBN.  Current
limits on the primordial element abundances are so tight
that even models with homogeneous degeneracy are tightly constrained.
Similarly, inhomogeous neutrino degeneracy does not allow
for a significant decrease in $\eta$, and such models tend
to give a neutrino energy density in conflict with other
cosmological limits.  It is quite impressive that
even with the radical model discussed here, the limits
on the primordial element abundances have become so tight
that a significantly lower value of $\eta$ cannot be achieved.
On the other hand, inhomogeneous neutrino degeneracy can increase
the upper bound on
$\eta$ to quite large values: up to $\eta = 1.0 \times 10^{-9}$
for the case of equal degeneracies in all three neutrinos,
and $\eta = 1.1 \times 10^{-8}$ if only $\nu_e$ is degenerate.
These correspond to $\Omega_b h^2 = 0.036$ and $0.40$, respectively.
 
The distributions of $\xi$ which produce these extreme
values for $\eta$
do not correspond to physically likely models.
The usefulness of our linear programming calculation is
that it allows us to establish upper and lower
bounds on $\eta$ for arbitrary distributions of $\xi$,
while at the same time giving the smallest value for $\rho_\nu^\prime/\rho_\nu$
corresponding to a given value of $\eta$.  Any other
distribution $f(\xi)$ is guaranteed to give values of $\eta$
which lie inside of our bounds.

This work does not exhaust the possible models with spatially-varying
$\xi$.  It is possible to use our methodology to investigate
models in which two or three neutrino degeneracies are independent.
In addition, if baryogenesis is related to the lepton number in
some way \cite{mcdonald,murayama}, then one would expect a correlation between $\eta$ and
$\xi$ at each point in space.  Given a specification for this correlation,
such models could also be examined within the framework we have outlined
here.  Less general but more physically realistic distributions
for $f(\xi)$ (e.g. a Gaussian distribution) could also be considered.

\acknowledgements

We thank A. Dolgov, G. Steigman and D. Weinberg for helpful discussions.
S.E.W. was supported at Ohio State under the NSF
Research Experience for Undergraduates (REU) program
(NSF PHY-9605064).
R.J.S. is supported by the Department of Energy
(DE-FG02-91ER40690).

%
%

\begin{table}
\caption{The nonzero values of $\xi_j$, along with the corresponding
$p_j$ and total $\rho_\nu^\prime/\rho_\nu$ for the case
where $\xi = \xi_e$, and $\xi_\mu = \xi_\tau = 0$.}
\begin{tabular}{|r|r|r|r|}
$\eta$ & $\xi_{j}$ & $p_j$ & $\rho_\nu^\prime/\rho_\nu$ \\ \hline
$3.0 \times 10^{-10}$ & $-0.8$ & 0.49 & 31 \\
 & 8 & 0.03 &  \\
 & 9 & 0.48 & \\ \hline
$1.1 \times 10^{-8}$ & $-37$ & $7.3 \times 10^{-4}$ & 4.3 \\
 & $-2$ & 0.23 & \\
 & 3 & 0.30 & \\
 & 4 & 0.47 & \\
\end{tabular}
\end{table}

\begin{table}
\caption{The nonzero values of $\xi_j$, along with the corresponding
$p_j$ and total $\rho_\nu^\prime/\rho_\nu$ for the case
where $\xi = \xi_e= \xi_\mu = \xi_\tau$.}
\begin{tabular}{|r|r|r|r|}
$\eta$ & $\xi_{j}$ & $p_j$ & $\rho_\nu^\prime/\rho_\nu$ \\ \hline
$3.1 \times 10^{-10}$ & $-0.7$ & 0.52 & 103 \\
 & 9 & 0.30 &  \\
 & 10 & 0.18 & \\ \hline
$1.0 \times 10^{-9}$ & $-25$ & $9.3 \times 10^{-5}$ & 3.2 \\
 & $-1$ & 0.37 & \\
 & 2 & 0.45 & \\
 & 3 & 0.18 &
\end{tabular}
\end{table}

\end{document}